\documentclass[prl,superscriptaddress,twocolumn,showpacs]{revtex4}
\usepackage{amsmath}
\usepackage{amssymb}
\usepackage{bm}
\usepackage{epsfig}
\usepackage{graphicx}
\usepackage{color}
\usepackage[english]{babel}

\def\be{\begin{equation}}
\def\ee{\end{equation}}
\def\bea{\begin{eqnarray}}
\def\eea{\end{eqnarray}}

\begin{document}

\newcount\timehh  \newcount\timemm
\timehh=\time \divide\timehh by 60
\timemm=\time
\count255=\timehh\multiply\count255 by -60 \advance\timemm by \count255

\title{Electron spin synchronization induced by optical nuclear magnetic resonance feedback}

\author{M. M. Glazov}
\affiliation{Ioffe Physical-Technical Institute RAS, 194021
  St.-Petersburg, Russia} 

\author{I. A. Yugova}
\affiliation{Institute of Physics, St. Petersburg State University,
  198504 St.-Petersburg, Russia} 

\author{Al. L. Efros}
\affiliation{Naval Research Laboratory, Washington DC 20375, USA}

\date{\today, Glazov-nuc-ae22.tex, printing time =
  \number\timehh\,:\,\ifnum\timemm<10 0\fi \number\timemm} 

%%% abstract 

\begin{abstract} 
We predict a new physical mechanism to explain the electron spin
precession frequency focusing effect recently observed in singly
charged quantum  
dots exposed to a periodic train of resonant circularly polarized
short optical pulses [A. Greilich et al, Science \textbf{317}, 1896
(2007), Ref. \cite{A.Greilich09282007}].   
We show that electron spin precession in an external magnetic field
and a field of nuclei creates a Knight field oscillating at the
frequency of the nuclear spin resonance.  This  
field drives the projection of the nuclear spin onto the magnetic field
to the value that makes the electron spin precession frequency a
multiple of the train cyclic  
repetition frequency, the condition at which the Knight field vanishes.
\end{abstract} 

\maketitle

An electron spin localized in a single quantum dot (QD) is a natural
qubit candidate for solid state quantum information processing
\cite{LossPRA98,WolfScience01,AwschalomCh02}.  
However, various optical or electrical control operations on an
electron spin in a QD affect the nuclear spin polarization (NSP),
which was observed as the Overhauser shift of  
the electron spin precession frequency in a magnetic field using
various pump-probe techniques
\cite{Gammon,KikkawaOld,Koppens06,Yamamoto}. The NSP is changed by  
electron-nuclear spin flip-flop processes resulting from Fermi contact
hyperfine interactions \cite{Abragam,dp74}.  Such processes, however,
are suppressed in a strong magnetic  
field because of an approximately three orders of magnitude mismatch
in energy between the electron and nuclear Zeeman splittings. The NSP
could be preserved on the time  
scale of hours \cite{NP}, unless special energy-conserving conditions
for electron spin-flip are reached \cite{ST}.  
%, such as, e.g., a singlet-triplet degeneracy in doubly charged QDs \cite{ST}. 
Consequently, various manipulations with an electron spin by optical
or electrical means become a main source of nuclear spin pumping, because
during the action of these time dependent  
perturbations spin-flip processes can occur without energy
conservation
\cite{A.Greilich09282007,
  carter:167403,Yamamoto,Nazarov,Levitov,Rashba,Korenev10}.   

One of the most remarkable demonstrations of such a phenomenon is the
nuclear induced frequency focusing (NIFF) effect that was discovered
in an ensemble of singly charged QDs  
under excitation by a periodic train of short resonant pulses of
circularly polarized light~\cite{A.Greilich09282007}. This experiment
showed that the nuclei change their  
polarization to values that allowed precession frequencies of all
electrons in the ensemble to satisfy the phase synchronization conditions
(PSC). These are the frequencies at which  
the Larmor precession period of electron spin is equal to an integer
fraction of the pump pulse repetition period
\cite{A.Greilich07212006}. Why does the NSP, which changes  
randomly under light excitation, reach the value allowing electrons to
satisfy the PSC? The authors of Ref.~\cite{A.Greilich09282007}
suggested a connection with suppression of  
nuclear spin dynamics in such dots. Indeed, the train
synchronizes the spin precession of electrons satisfying PSC and makes
them optically passive at the moment of pulse  
arrival. This significantly slows down the light-stimulated random
dynamics of the NSP in these QDs,
leading to the %NIFF and  
accumulation of electron spins satisfying the  
PSC~\cite{A.Greilich09282007}.

In this Letter we demonstrate that the NIFF could be the result of the
Knight field feedback-stimulated nuclear magnetic resonance (NMR). Our
calculations treat the electron  
spin and NSP as classical vectors precessing around each other and an
external magnetic field, and show that the NSP increases its
projection onto the magnetic field %direction  
monotonically with time therefore modifying the electron spin precession frequency. When the electron spin precession frequency has reached the PSC, the time-averaged electron 
spin polarization generated by the train and the corresponding Knight
field causing the NSP modifications vanish. The suggested
mechanism should result in much faster frequency focusing than 
%the NIFF 
that
connected with random fluctuations of the
NSP~\cite{A.Greilich09282007,carter:167403,Korenev10}

We consider a singly negatively charged QD exposed to the train of circularly polarized pump pulses propagating along the structure growth axis $z$, arriving at the QD with the 
repetition period $T_R$, and also to a transverse magnetic field $\bm B\parallel \bm e_x$, where $\bm e_x$ is the unit vector along $x$-axis (see inset in Fig.~\ref{electron}). 
It is 
assumed that the optical transition involves the excitation of a singlet $X^-$ trion with the hole spin projection on the growth axis being $\pm 3/2$ for $\sigma^+$ and 
$\sigma^-$ 
pump pulses, respectively. The pulse duration $\tau_p$ is short as compared with the spin precession period in the external magnetic field and as compared with the photocreated 
trion lifetime. The optical selection rules are therefore the same as in the absence of a magnetic field. In the interval between the optical pulses, the electron spin interacts 
with the NSP, $\bm m=\sum_i \bm I_{i}$  where $\bm I_i$ are the
nuclear spins and the sum is over a mesoscopic number ($N\sim 10^5$)
of nuclear spins. At equilibrium, in the studied magnetic fields, nuclei are
  practically 
unpolarized: they are randomly oriented and  the NSP magnitude is
controlled by random fluctuations 
of nuclear spin directions $|m| \sim \sqrt{N} \sim 3\times 10^2$.
To describe this electron-nuclei interaction we 
treat the electron spin polarization, $\bm S$, and $\bm m$ as classical vectors \cite{MerkulovPRB02} and adopt the box model~\cite{RS1983,Kozlov2007,PhysRevB.76.045312} in 
which the interaction between electrons and nuclear spins does not depend on their positions. These approximations lead to the following equations for $\bm S$ and $\bm m$ in the 
interval between the optical pulses, $(n-1)T_R \leqslant t < nT_R$, where $n$ is the pulse number \cite{PhysRevB.81.115107}:
\begin{subequations}\label{precess}
\bea
\label{electron}
\frac{\mathrm d\bm S}{\mathrm dt}& = &[(\bm \Omega +\alpha \bm m(t)) \times \bm S(t)], \\
\label{nuclear}
\frac{\mathrm d\bm m}{\mathrm dt} &=& [(\bm \omega + \alpha \bm S(t)) \times \bm m(t)].
\eea
\end{subequations}
Here $\bm \Omega=\Omega \bm e_x$ and $\bm \omega = \omega \bm e_x$ are the electron and nuclear spin precession frequencies in an external field, and $\alpha$ is the hyperfine 
coupling constant between the electron and nuclear spins in the QD. The difference of electron and nuclear magnetic moments gives $\omega/\Omega \sim 10^{-3}$. The electron and 
nuclear spins in Eqs.~\eqref{precess} are coupled via an Overhauser field of NSP fluctuation acting on the electron, $\alpha \bm m$, and a Knight field of the electron spin 
acting on the nuclei, $\alpha \bm S$. We neglect completely a slow nuclear spin relaxation connected with dipole-dipole interactions between nuclei in Eq.~\eqref{nuclear}.
 
The dynamics of the electron and nuclear spins in the QD have several
very different time-scales. Under experimental
  conditions~\cite{A.Greilich09282007} following inequalities hold:
\[
\frac{2\pi}{\Omega} \ll \frac{2\pi}{\alpha m} \lesssim T_R \ll
\frac{2\pi}{\omega} \ll
\frac{2\pi}{\alpha}.
\]
These inequalities mean (i) that  electron spin dynamics in the
interval between pulses occurs in the permanent  field of the frozen
fluctuation of NSP; 
and (ii) that  the dynamics of NSP is 
controlled only by the electron spin polarization averaged over the
pulse repetition period: 
\begin{equation} \bm S_0 = \frac{1}{T_R} \int_{(n-1)T_R}^{nT_R}\bm S(t)  \mathrm dt~. 
\end{equation} 
Straightforward calculation shows that 
\begin{multline}\label{s0} \bm S_0 = \bm n (\bm n \cdot \bm S^{(a)} ) + \frac{\bm S^{(a)} - \bm n (\bm n\cdot \bm S^{(a)} )}{\Omega_{\rm eff} T_R} 
\sin{(\Omega_{\rm eff} T_R)} \\ + \frac{[\bm S^{(a)} - \bm n (\bm n
  \cdot \bm S^{(a)} )]\times \bm n}{\Omega_{\rm eff} T_R} [1-\cos{(\Omega_{\rm eff} T_R)}], 
\end{multline} 
where $\bm S^{(a)}$ is the electron spin polarization right after the excitation pulse, $\bm n = (\bm \Omega + \alpha \bm m)/\Omega_{\rm eff}$ is a unit vector along the 
effective field and $\Omega_{\rm eff} = |\bm \Omega + \alpha \bm m| \approx \Omega + \alpha m_x$. One can see from Eq. \eqref{s0} that the average electron spin 
polarization $\bm S_0$ transverse to $\bm n$ vanishes when
$\Omega_{\rm eff}$ satisfies the PSC: $\Omega_{\rm eff} T_R = 2\pi K,$
with $K=1,2, \ldots$ {The longitudinal component does not
  vanish at the PSC due to a small deviation of $\bm n$ from the
  magnetic field direction caused by the nuclear field.  As we show
  below, the transverse components of an electron spin are required
  for the NSP modification. As a result,  the electron spin  does not affect the nuclei if the PSC is fulfilled.}  If the PSC
is not satisfied, however, the weak Knight field, $\alpha \bm S_0$,
modifies the NSP and drives its projection, $m_x(t)$, to the value that allows
$\Omega_{\rm eff}$ to satisfy the PSC.

\begin{figure}[hptb]
\includegraphics[width=0.75\linewidth]{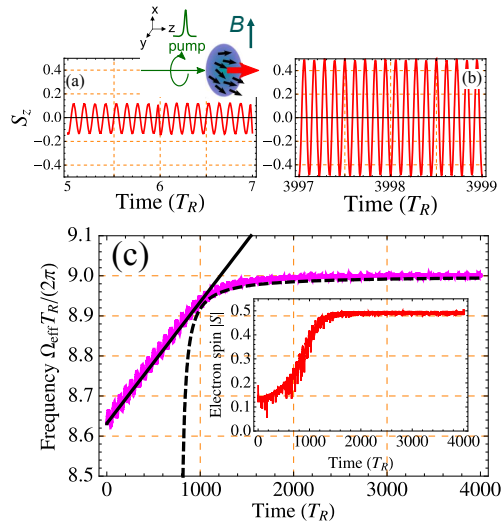}
\caption{Time dependence of $z$ component of the electron spin polarization $S_z$ calculated after the train initiation (a), and after $\sim 4000$ repetition periods 
of the train (b). Panel (c) shows electron spin precession frequency calculated numerically (magenta) and analytically from Eq.~\eqref{mx1} (black solid) and 
Eq.~\eqref{mx21} (black dashed) curves. Inset to panel (a) illustrates geometry of a single QD excitation and shows a pump pulse and an electron (red) and nuclear 
(black) spins. Inset to panel (c) shows the absolute value of electron spin as a function of time. Calculations were conducted for $\alpha=0.4$, $m=23.5$, which 
corresponds to approximately $2200$ nuclei with spin $I=1/2$,
$\Theta=2\pi/3$, $\Omega T_R/(2\pi) = 8.5$, and $\omega=\Omega/500$.}\label{fig:electron}
\end{figure}

To describe this effect we need to complement Eq.~\eqref{precess}, which describes the electron-nuclear spin dynamics in the interval between pulses, by the relationship 
between the electron spin polarization before, $\bm S^{(b)}$, and after, $\bm S^{(a)}$, the pump pulse, which for resonant excitation read~\cite{yugova09}
\be
\label{eq16}
S_z^{(a)} = \frac{Q^2-1}{4}+S_z^{(b)}\frac{Q^2+1}{2}, S_x^{(a)}=QS_x^{(b)}, S_y^{(a)} = QS_y^{(b)},
\ee 
where $Q=\cos{\Theta/2}$ and $1-Q^2$ is the probability of trion creation by the short circularly polarized pulse with area $\Theta$. 
Numerical integration of Eqs.~\eqref{precess}, which uses Eq.~\eqref{eq16}, clearly demonstrates the NIFF effect as one can see in Fig.~\ref{fig:electron}.  
Calculations were conducted for the electron spin precession
frequency, which does not satisfy the PSC: $\Omega T_R/(2\pi) = 8.5$,
and an initial condition for the NSP,  
which was selected as $\bm m \parallel \bm e_z$. We exaggerated the
value of $\alpha$ and reduced the number of nuclei from a typical
value in a QD $N\sim 10^5$ down to $N\sim 2\times 10^3$ to conduct
numerically accurate calculations within reasonable computational
time. Otherwise the  difference in the characteristic times of
electron and nuclei spin dynamics requires carrying out calculations
on a timescale covering nine orders of magnitude.

Figure~\ref{fig:electron}(a) shows the temporal dynamics of the electron spin $z$-component for the $6^{\rm th}$ and $7^{\rm th}$ repetition periods where the electron 
spin dynamics is already stationary~\cite{A.Greilich07212006} but the nuclear effects have not come into play. Panel (b) shows those dynamics for the $3998^{\rm th}$ and 
$3999^{\rm th}$ periods when the nuclear spin precession had already taken place. One can see that the slow nuclear spin dynamics changes qualitatively the character of 
electron spin precession in this time interval:  the amplitude of electron spin polarization is strongly enhanced and reaches its maximum value 1/2, see inset in 
Fig.~\ref{fig:electron}(c).  The effect is connected with the temporal dynamics of the electron effective spin precession frequency shown in Fig.~\ref{fig:electron}(c). 
Apart from the oscillations of frequency $\omega$ related to the NSP precession, $\Omega_{\rm eff}$ initially grows linearly in time and then saturates at the 
multiple of $2\pi/T_R$.  The periodic train of short pulses synchronizes the electron spin precession in the QD where $\Omega_{\rm eff}$ satisfies the PSC, leading 
to complete polarization of electron as seen in Fig. \ref{fig:electron}(b).

\begin{figure}[hptb]
\includegraphics[width=0.9\linewidth]{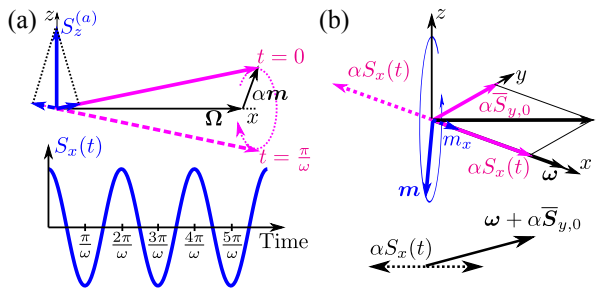}
\caption{(a) Schematic illustration of electron spin precession in
  quasi-static field $\bm \Omega+ \alpha\bm m(t)$ (top) and temporal
  dependence of $S_x$ (bottom). (b) Geometry of NMR induced by steady
  state $\alpha \overline{S}_{y,0}$, and alternating,
  $\alpha S_{x}(t)$, Knight fields. Bottom panel shows
    static and oscillating fields in the $(xy)$ plane.} \label{fig:illustr}
\end{figure}

To understand physical mechanism responsible for NIFF demonstrated in Fig.~\ref{fig:electron}, let us first consider the effect of the nuclear spin precession on the 
electron spin dynamics. 
Since nuclear spin precession is slow as compared with
  electron spin precession and pump pulse repetition periods,
one can 
treat the electron spin dynamics in the interval between pulses as a
precession in the static field $\bm \Omega + \alpha \bm m$ (see
Fig.~\ref{fig:illustr}a). 
Using procedure from Ref.~\cite{yugova09} and
  Eqs.~\eqref{s0}, \eqref{eq16} we derive a steady state 
  (on the timescale of $T_R$) value of
  $S_{x}$ exposed  to the train of optical pulses:
\begin{equation}
\label{sxt}
S_x \equiv S_{x,0}(t) = \alpha m_z(t) C_x/\Omega,
\end{equation}
\[C_x = -\frac{2Q\sin^2{\left(\Omega_{\rm eff} T_R/2\right)}+(Q-1)^2/2}{(Q-1)^2+2(Q+1)\sin^2{\left(\Omega_{\rm eff} T_R/2\right)}},
\]

One can see from Eq.~\eqref{sxt} that $S_x(t)$ oscillates slowly with
the NSP precession frequency $\omega$. This occurs because the
electron spin precession axis (see  
Fig.~\ref{fig:illustr}a) is tilted from the $x$ axis in the $(xz)$
plane by the small angle $\alpha m_z(t)/\Omega$. The precession leads
to a non-zero $S_x$-projection  
of the electron spin, which value is proportional to the tilt
angle oscillating  at the frequency $\omega$. The same geometrical
arguments show that $S_{0,y}$ and $S_{0,z}$ are the sum of the time
independent terms $\overline{S}_{0,y}$ and $\overline{S}_{0,z}$ and
small terms oscillating at $2\omega$ that can be neglected. As a
result, the NSP, which precesses around the static field $\bm \omega
+\alpha \overline{\bm S}_0$ slightly tilted from the $x$ axis 
 experiences 
the alternating Knight field $\alpha S_x(t)$ (see
Fig.~\ref{fig:illustr}b). Since $S_x(t)$ oscillates with $\omega$ it
drives the NMR and leads to slow modification of  
$m_x$, as shown below. 

To describe the time dependence of $m_x(t)$ we need to take into
account that the NMR driving field, $\alpha S_x(t)$, is almost
parallel to the static field.  At first, 
it creates a time-independent shift of the NSP, $\bar m_z$, along the $z$ axis. Indeed, in the first approximation on $\alpha$: 
\begin{equation} \label{mzt} 
m_z(t)=m_\perp\cos{\left[\omega t + \int_0^t \alpha S_x(t') \mathrm d t'\right]} 
\end{equation} 
where $m_\perp=\sqrt{m^2-m_x^2}$ is the perpendicular component of the NSP. In the same approximation on $\alpha$, Eq.~\eqref{mzt} can be rewritten as: $m_z(t)\approx 
m_\perp\cos{\omega t} + \bar m_z$, where $\bar m_z=-\alpha^2
m_\perp^2C_x/(2\omega\Omega)$. The analogous calculation shows
  that $\bar m_y=0$.

Secondly, the NMR is caused only by the component of the alternating field perpendicular to the NSP precession axis, which is equal to $\alpha(\alpha \bm 
S_0/\omega)S_x(t)$. 
Averaging the $x$-component of Eq.~\eqref{nuclear}: $\mathrm d
  m_x/\mathrm dt = \alpha (S_y m_z - S_z m_y)$ over a sufficiently
  long temporal interval $\Delta T \gg 
  1/\omega\gg 1/\Omega$  we obtain the standard NMR expression:
\begin{equation}
\label{mx}
\frac{\mathrm d m_x}{\mathrm dt} = \alpha \overline{S}_{y,0} \bar m_z =-  \frac{\alpha^3 \overline{S}_{y,0} C_x m_\perp^2}{2\omega \Omega}, 
\end{equation}
 where the averaged $m_y$: $\bar m_y=0$.
Generally, the right hand side of Eq.~\eqref{mx} depends on $m_x$ via $S_{y,0}$ and $C_x$ dependence on $\Omega_{\rm eff}$. One can neglect this dependence if 
$\Omega_{\rm eff}$ is not very close to the PSC. In this case we obtain for $m_x(t)$:
\begin{equation}
\label{mx1}
\frac{m_x(t)}{m_\perp(0)} \approx \frac{t}{\tau_{\rm nf}} , \quad \frac{1}{\tau_{\rm nf}} = -\frac{\alpha^3m_\perp(0)}{2\omega\Omega} \overline{S}_{y,0} C_x,
\end{equation}
where $m_\perp(0)$ is  the initial value of the perpendicular
component of the NSP. One can see from Eq. \eqref{mx1} that $m_x(t)$,
and consequently $\Omega_{\rm eff}$, grow linearly with time. The
analytical dependence $\Omega_{\rm eff}(t)$ shown by the  
solid line in Fig.~\ref{fig:electron}(c) is in good agreement with
results of the numerical calculations.

In the case that $\Omega_{\rm eff}$ is close to the phase
  synchronization condition, which is fulfilled if $m_x=m_x^{\rm PSC}$, one can rewrite Eq. \eqref{mx} using Eq.~\eqref{s0} as: 
\begin{equation}
\label{mx2}
\frac{\mathrm d m_x}{\mathrm dt} = \frac{\left(m_x-m_x^{\rm PSC}\right)^2}{m\tau_{\rm nf}'}, 
\end{equation}
where 
\begin{equation}
\label{taunf1}
\frac{1}{\tau_{\rm nf}'} = \frac{\alpha^5 m T_R}{16\omega\Omega^2}\frac{1+Q}{1-Q}\left[m^2-(m_x^{\rm PSC})^2\right].
\end{equation}
The dynamics of $m_x(t)$ in this case is described by
\begin{equation}
\label{mx21}
m_x(t) = m_x^{\rm PSC} - m\frac{\tau_{\rm nf}'}{t-t_0},
\end{equation}
where $t_0$ is an arbitrary constant, chosen to merge the time dependencies given by Eqs.~\eqref{mx1} and \eqref{mx21} at $t\sim \tau_{\rm nf}$. Corresponding long-time 
asymptote of $\Omega_{\rm eff}$ is plotted in Fig.~\ref{fig:electron}(c) by a dashed line.

Although $\mathrm dm_x/\mathrm dt=0$ at $m_x = m_x^{\rm PSC}$ these $m_x$'s are only the saddle points in the $m_x$ time dependence. For positive $\tau_{\rm nf}'$ 
the points are stable if $m_x<m_x^{\rm PSC}$ and unstable
otherwise. This means that $m_x$  returns  to $m_x^{\rm
    PSC}$  only  if   its  fluctuation  $\delta =m_x- m_x^{\rm
    PSC}<0$. If $\delta>0$, the fluctuation 
causes the deterministic growth of  $m_x$  until the next PSC with larger $m_x$
  is met.   Including a weak nuclear spin relaxation in
Eq. \eqref{nuclear} gives two $m_x=m_x^{\rm  
PSC}\pm \sqrt{m_x^{\rm PSC} m \tau_{\rm nf'}/T_1}$, at which $\mathrm
dm_x/\mathrm dt=0$, where $T_1\gg \tau_{\rm nf'}$ is the nuclear spin
relaxation time. One of these  
solutions, $m_x=m_x^{\rm PSC}- \sqrt{m_x^{\rm PSC}m \tau_{\rm
    nf'}/T_1}$, is a stable point of $m_x(t)$. A switch of the light
polarization from $\sigma^+$ to  
$\sigma^-$ does not change the direction of the $m_x$ growth, as can
be seen from Eqs. \eqref{mx1} and \eqref{mx21}. 

\begin{figure}[hptb]
\includegraphics[width=0.8\linewidth]{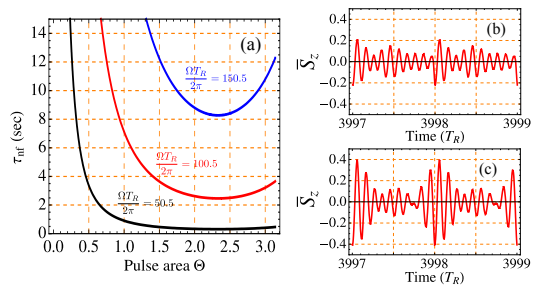}
\caption{(a) Dependence of the NIFF time, $\tau_{\rm nf}$, on the pulse area, $\Theta$. The three curves were calculated at the magnetic fields which give the following electron spin precession frequencies: $\Omega T_R/(2\pi)=50.5$ (black), $100.5$ (red) and $150.5$ (blue). The parameters used:  $\alpha=5\times 10^{6}$~sec$^{-1}$, $T_R=13$~ns, $\omega/\Omega=10^{-3}$ and $m=126$, which could be created by $6\times 10^4$ nuclei with spin $1/2$, were selected to keep the calculations relevant to  Refs.~\cite{A.Greilich07212006,A.Greilich09282007}. (b) and (c) Time dependence of the average $z$ component of the electron spin $\overline{S}_z(t)$ during  $3997T_R <t< 3999T_R$ time interval. The  averaging was conducted over 25 initial directions of NSP $\bm m$. Panel (b) is calculated for the frozen nuclear fluctuation ($\alpha=\omega=0$ in Eq.~\eqref{nuclear}), and panel (c) is calculated taking the nuclear spin precession into account ($\alpha=0.4$, $\omega=\Omega/500$). Other parameters are the same as in Fig.~\ref{fig:electron}.} \label{fig:taunf}
\end{figure}

Figure~\ref{fig:taunf} shows the dependence of the NIFF time, $\tau_{\rm nf}$, defined Eq.~\eqref{mx1} on the pump pulse area, $\Theta$. One can see that $\tau_{\rm nf}$ 
becomes extremely long for $\Theta \ll 1$ because electron spin orientation is inefficient and the averaged electron spin $\bm S_0$ is very small under these conditions. 
Growth of $\Theta$ increases $\bm S_0$ and consequently the Knight field, $\alpha\bm S_0$, which in turn shortens $\tau_{\rm nf}$.  Further increase of $\tau_{\rm nf}$ 
with $\Theta$ seen in Fig.~\ref{fig:taunf} is connected with the
periodic dependence of $\bm S_0$ on $\Theta$. The frequency
  focusing time $\tau_{\rm nf} \propto \Omega^2\omega$ 
increases significantly with a magnetic field. This explains the rapid increase of $\tau_{\rm nf}$ with $\Omega$ seen in Fig.~\ref{fig:taunf}.

We have considered electron-nuclear spin dynamics in a single QD with a certain initial orientation of NSP. To describe a QD ensemble we average over different initial 
orientations of the NSP. The time dependence of the average $z$-component of the electron spin $\overline{S}_z(t)$ is shown in Fig.~\ref{fig:taunf}(b),(c).  The initial 
NSP orientations $\bm m(0)$ were chosen to be isotropically distributed, with the magnitude $m(0)=m=23.5$ used to describe the spin dynamics of a single QD in 
Fig.~\ref{fig:electron}.  Figure~\ref{fig:taunf}(b) shows the electron spin dynamics in the absence of nuclear spin dynamics, which is simulated using $\alpha=0$ 
and $\omega=0$ in Eq.~\eqref{nuclear}. One can see that $\overline{S}_z(t)$ partially decays between the pump pulses and the phase of spin beats jumps at each repetition 
period.  Comparison with Fig.~\ref{fig:electron}(a) shows that the amplitude of $\overline{S}_z(t)$ after 4000 repetition periods is the same as at the initial 
precession stage. The nuclear spin precession in the external magnetic field and in the Knight field tunes up the electron spin precession frequency and leads to a very 
pronounced mode-locking of electron spin coherence seen in Fig.~\ref{fig:taunf}(c). One can see a clear rise of $\overline{S}_z(t)$ before pulse arrival. The NIFF effect 
also significantly increases the $\overline{S}_z(t)$ amplitude relative to those shown in Fig.~\ref{fig:taunf}(b). For the parameters used in this calculation $\alpha m 
T_R/(2\pi)\approx 1.5$ the main contribution to $\overline{S}_z(t)$
comes from the two $\Omega_{\rm eff}$'s satisfying the PSC
  $\Omega_{\rm eff}T_R/2\pi=  8\mbox{ and }9$.

We note that if the NIFF effect is a consequence of random fluctuations of the NSP as suggested in Refs.~\cite{A.Greilich09282007,carter:167403},  the rate of this 
process  can be estimated as $\gamma_{\rm n} \sim
(1-Q)\alpha^2/(\Omega^2 T_R)$. The current model leads to a much
faster NIFF in the QD ensemble studied in
Ref. \cite{A.Greilich09282007} because $1/(\gamma_{\rm n}\tau_{\rm
  nf}) \sim \alpha m/\omega \sim 10$ in these
experiments. %\addMisha{We have carried out also the modeling of NIFF in the graded box approach~\cite{petrov} considering three different groups of nuclear spins with three different coupling constants. We did not observed any suppression of the average rate of the NIFF. For some initial configurations of the nuclei spins the frequency focusing time increases by about a factor two, but for majority  the configurations focusing goes much faster  The acceleration of the NIFF effect may be connected with the fact that in the grated box an electron predominantly interacts with nuclei at the QD center due to the largest coupling constant proportional to the electron density. The same nuclei are responsible for the feedback strength.  This leads to the faster  NIFF because the effective coupling constant responsible for the focusing   is larger than the average one. } 

In summary, we have suggested a new physical mechanism of the NIFF
effect for the electron spin precession \cite{A.Greilich09282007}.
This mechanism leads to a  
monotonic shift of the electron spin precession frequency with time
and allows this frequency to reach phase synchronization condition with the train repetition
period much faster than in the case  
when the NIFF is a consequence of random fluctuations of the
nuclear spins as
was suggested earlier. Further experimental studies of the NIFF time
dependence on a magnetic field and a pulse area should provide 
evidence for the suggested mechanism.

\emph{Acknowledgements.} 
Authors thank A. Braker, I.V. Ignatiev, E.L. Ivchenko, E.I. Rashba for
valuable discussions and RFBR, ``Dynasty'' Foundation---ICFPM,
Alexander von Humboldt Foundation and the Office of Naval Research for
support.

\end{document}